\documentclass[twocolumn,showpacs,preprintnumbers,amsmath,nofootinbib,amssymb]{revtex4}
\input psfig.sty
\def \be {\begin{equation}}
\def \ee {\end{equation}}
\def \bea {\begin{eqnarray}}
\def \eea {\end{eqnarray}}

\usepackage{graphicx}
\usepackage{dcolumn}
\usepackage{bm}

\begin{document}

\title{Cosmological bounds on oscillating dark energy models}

\author{Deepak Jain$^1${\footnote{djain@ddu.du.ac.in}}}

\author{Abha Dev$^2$}

\author{J. S. Alcaniz$^3${\footnote{alcaniz@on.br}}}

\affiliation{$^1$Deen Dayal Upadhyaya College, University of Delhi, New Delhi 110015, India}

\affiliation{$^2$Miranda House, University of Delhi, Delhi 110007, India}

\affiliation{$^3$Departamento de Astronomia, Observat\'orio Nacional,
20921-400 Rio de Janeiro - RJ, Brasil}

\date{September 26, 2007}

\begin{abstract}

We study some cosmological constraints on the two phenomenological models of oscillating dark energy. In these scenarios, the equation of state of dark energy varies periodically and may provide a way to unify the early acceleration (inflation) and the late time acceleration of the universe. These models give also an effective way to tackle the so-called cosmic coincidence problem. We examine observational constraints on the oscillating models from the latest observational data including the \emph{gold} sample of 182 type Ia supernovae, the CMB shift parameter $R$ and the BAO measurements from the Sloan Digital Sky Survey.

\end{abstract}


\maketitle

\section{Introduction}

There are growing observational evidence that the cosmic expansion is speeding up and that the universe underwent a recent transition from a decelerating to accelerating phase. This late-time acceleration in turn poses a major theoretical challenge in cosmology and has been evidenced by a number of independent observational results, which includes distance measurements to Type Ia Supernova (SNe Ia) \cite{sn1,sn2,sn3}, current Cosmic Microwave Background (CMB) anisotropies measurements \cite{cmb1,cmb2}, and data of the Large-Scale Structure (LSS) in the Universe \cite{lss}. These observations are often 
explained by introducing a new hypothetical energy component, usually referred to as dark energy. The nature of such dark energy component constitutes a
completely open question nowadays and the only facts about this componenet are  that (1) it has a negative pressure (2) its energy density is of the order of the critical
density, $\sim 10^{-29} \rm{g/cm^3}$) and (3) it is unclustered. Deciding between the many possible sources of acceleration will be one of the  major thrusts in observational cosmology in the next decades (for recent reviews on this topic, see \cite{revde}).

Apart from their own consistency (theoretical and observational) problems, almost all the current candidates for dark energy also face the so-called ``coincidence problem", i.e.,  why the dark energy density is approximately equal to the matter density at this moment of cosmic history. The spectrum of the possible answers to
this question includes anthropic rationalizations \cite{st}, tracking fields \cite{tf}, as well as cyclic cosmologies \cite{cc}. In this regard, an interesting question put forward in Ref. \cite{tf1} concerns to the possibility of a periodical accelerating expansion of the Universe over its past evolution. If this is so, then the
fact that the Universe is accelerating today would not be surprising. In fact, it would be reasonable to expect it to accelerate today. Following this reasoning, several interesting oscillating dark energy models have been discussed in literature \cite{os}. Although most of these scenarios are motivated from purely phenomenological
reasons, in some of them a periodical acceleration can emerge from the evolution of a single scalar field in a potential with oscillatory and exponential behavior \cite{bar} and also constitutes a new kind of approach to the idea of quintessential inflation.

In this \emph{Letter}, motivated by a possible amelioration of the coincidence problem, we aim at testing the viability of two different parameterizations for oscillating dark energy in light of some of the most recent observational data. We focus our attention on the equation of state (EoS) parameterizations  $\omega(a) = - \cos( b\, \ln a)$ (P1) and $\omega(a) = \omega_0 - A\sin(B\ln a)$ ({P2), as discussed in Refs. \cite{un} and \cite{lin}, respectively. For our observational analysis, we use the latest Supernova \emph{gold} sample set of 182 SNe Ia \cite{sn3} along with the current estimate of the CMB shift parameter \cite{cmb2}, and the Sloan Digital Sky Survey (SDSS) measurement of the baryon acoustic oscillation (BAO) peak \cite{sdss}. Motivated by inflation (and as evidenced by the combination of the position of the first acoustic peak of the CMB power spectrum and the current value of the Hubble parameter), we assume spatial flatness.  

The plan of the paper is as follows. In Section II, we  describe the cosmological scenarios arising from the parameterizations above and derive their basic equations. The observational analysis testing the viability of these oscillating dark energy models are discussed in Section III. We end the paper by summarizing our results in Section IV.

\section{Models}

\noindent In this work, we are particularly interested in two specific
parameterizations for the oscillatory behavior of the dark energy EoS:

\be
\omega(a) = - \cos( b\, \ln a)\; , \quad \quad \quad \quad \quad \rm{(P1)}
\ee
and
\be
\omega(a) = \omega_0 - A\sin(B\ln a)\; , \quad \quad \quad \rm{(P2)}
\ee 
where $a$ is the cosmological scale factor.

In the first parameterization (P1) \cite{un}, the expansion of the so-called undulant universe is characterized by alternating periods of acceleration and deceleration with the dimensionless parameter $b$ controlling the frequency of the accelerating epochs. Note that in the limit of small values of $ b$, the above EoS approaches the cosmological constant, $ \omega \sim -1$. The sinusoidal parameterization (P2) \cite{lin} follows the usual form of varying EoS parameterizations and has $\omega_0$ as the center of the range over which $\omega$ oscillates with the parameters $A$ and $B$ standing for the amplitude of oscillations $A$ and the frequency of oscillations, respectively. Note also that as the frequency of oscillations $B$ increases,  observational tests which involve integrals of equation of
state may fail to distinguish the oscillatory model from a model with a constant equation of state. In the subsequent analysis we assume that $w(z) \ge -1$ which requires $\omega_0 - A \ge -1$.  

Since Eqs. (1) and (2) represent separately conserved components, it is straightforward to show from the energy conservation law [$\dot{\rho}_j = -3\dot{a}/a(\rho_j + p_j )$] that the ratio $f_j = \rho_j/\rho_{j0}$ for (P1) and (P2) evolves, respectively, as 
\be
f_1 = a^{-3}\exp\left [\,({3\over b})\,\, \,{\sin} ( b\, \ln a)\right ]\;,
\ee 
and
\be
f_2 = a^{-3(1+\omega_0)}\exp\left\{\left(\frac{3A}{B}\right)\left[1-\cos(B\ln a)\right]\right\}
\ee 
with the subscript $j = 1,2$. Finally, the Hubble expansion for the oscillating dark energy models described above is given by 
\be
\label{Hz}
H(z; \mathbf{s}) = H_0 E(z; \mathbf{s}) = H_0\left[\Omega_m (1 + z)^{3} + \Omega_j f_j \right]^{1/2}.
\ee
where $\mathbf{s} \equiv (\Omega_{m}, w_0, b, A, B)$ is the complete set of parameters.

\section {Testing oscillating EoS}

In order to test the viability of the above scenarios and place limits on parameterizations P1 and P2 we use three of the most recent sets of observations currently available in the literature.

\subsection{SNe Ia}

The supernova data we use are the ``gold''  set of 182 SNe Ia recently published by Riess et. al  (2006) \cite{sn3}. This set  includes of  119 data points from the
previous sample analyzed by Riess et al. (2004) \cite{sn1}, 16 new  SNe Ia discovered by Hubble Space Telescope ( HST) in the range $0.46 < z < 1.39$ \cite{sn1} and 47 points from the the first year  release of SNLS dataset $( 0.25 < z < 0.96)$ \cite {sn2}.

The predicted distance modulus for a supernova at redshift $z$, given the set of parameters $\mathbf{s}$, is
\begin{equation} \label{dm}
\mu_p(z|\mathbf{s}) = m - M = 5\,\mbox{log} d_L + 25,
\end{equation}
where $m$ and $M$ are, respectively, the apparent and absolute magnitudes, and $d_L$ stands for the luminosity distance (in units of megaparsecs),
\begin{equation}
d_L = c(1 + z)\int_{x'}^{1} {dx
\over x^{2}{H}(x;\mathbf{s})},
\end{equation}
\noindent with $x' = (1 + z)^{-1}$ being a convenient integration variable, and ${H}(x; \mathbf{s})$ the expression given by Eq. (\ref{Hz}). We estimated the best fit to the set of parameters $\mathbf{s}$ by using a $\chi^{2}$
statistics, with
\begin{equation}
\chi^{2}_{\mathrm SNe} = \sum_{i=1}^{N}{\frac{\left[\mu_p^{i}(z|\mathbf{s}) -
\mu_o^{i}(z)\right]^{2}}{\sigma_i^{2}}},
\end{equation}
where $N = 182$, $\mu_p^{i}(z|\mathbf{s})$ is given by Eq. (\ref{dm}), $\mu_o^{i}(z)$ is the extinction corrected distance modulus for a given SNe Ia at $z_i$, and $\sigma_i$ is the uncertainty in the individual distance moduli.

\subsection{Baryon acoustic oscillations}

The acoustic peaks in the CMB anisotropy power spectrum is an efficient way for determining cosmological parameters. Because the acoustic oscillations in the
relativistic plasma of the early universe will also be imprinted on to the late-time power spectrum of the non relativistic matter, the acoustic signatures in the
large-scale clustering of galaxies yield additional tests for cosmology. In particular, the characteristic and reasonably sharp length scale measured at a wide range of redshifts provides an estimate of the distance-redshift relation, which is a geometric complement to the usual luminosity-distance from SNe Ia. Using a large spectroscopic sample of 46,748 luminous-red galaxies covering 3816 square degrees out to a redshift of $z=0.47$ from the SDSS, Eisenstein {\it et al.} \cite{sdss} have successfully found the peaks. The SDSS BAO measurement provides ${\cal{A}}_{\mathrm{obs}} = 0.469(n_S/0.98)^{-0.35} \pm 0.017$, with ${\cal{A}}$ defined as
\begin{equation}
{\cal{A}} \equiv
\frac{\Omega_{\rm{m}}^{1/2}}{z_{\rm{*}}}\left[z_{\rm{*}}
  \frac{\Gamma^{2}(z_{\rm{*}};\mathbf{s})}{{\rm{E}}(z_{\rm{*}};\mathbf{s})}\right]^{1/3}, 
\end{equation}
\noindent where $z_{\rm{*}} = 0.35$, $\Gamma(z_{\rm{*}}) =
\int_0^{z_{\rm{*}}}dz/E(z_{\rm{*}})$ 
is the dimensionless comoving distance to $z_{\rm{*}}$,
${\rm{E}}(z_{\rm{*}};
\mathbf{s})$ is given by Eq. (\ref{Hz}), and we  take the scalar spectral index $n_S = 0.95$, as given in Ref. \cite{cmb2}.

\subsection{CMB shift parameter}

The shift parameter  ${\mathcal{R}}$, which determines the whole shift of the CMB angular power spectrum, is given by \cite{b}
\begin{equation}
{\mathcal{R}}={\sqrt{\Omega_{m}}}\int_0^{z_{\mathrm ls}}\frac{dz}{E(z)}\;, 
\label{cmb}
\end{equation}
where $z_{\mathrm ls} = 1089$ is the redshift of the last scattering surface and the current value of the above quantity is ${\mathcal R}_{\mathrm{obs}} = 1.70\pm 0.03$ \cite{yw}. To perform our statistical analysis we use a $\chi^2$ statistics defined as $\chi^2_{\rm{T}}=\chi^2_{\mathrm SNe} + \chi^2_{\mathrm  BAO} + \chi^2_{\mathrm CMB}$, where $\chi^2_{\mathrm BAO}= {{(A - A_{\mathrm{obs}})}^2}/{\sigma_{A}^2}$ and $\chi^2_{\mathrm CMB}={({\mathcal R} - {\mathcal R}_{\mathrm{obs}})^2/{\sigma_{\mathcal R}}^2}$(for more  details on the statistical analysis we refer the reader to Ref.~\cite{refs}).

\begin{figure*}[]
{
\setlength{\unitlength}{0.240900pt}
\ifx\plotpoint\undefined\newsavebox{\plotpoint}\fi
\sbox{\plotpoint}{\rule[-0.200pt]{0.400pt}{0.400pt}}%
\begin{picture}(1500,900)(0,0)
\font\gnuplot=cmr10 at 10pt
\gnuplot
\sbox{\plotpoint}{\rule[-0.200pt]{0.400pt}{0.400pt}}%
\put(201.0,123.0){\rule[-0.200pt]{4.818pt}{0.400pt}}
\put(181,123){\makebox(0,0)[r]{ 0}}
\put(1419.0,123.0){\rule[-0.200pt]{4.818pt}{0.400pt}}
\put(201.0,270.0){\rule[-0.200pt]{4.818pt}{0.400pt}}
\put(181,270){\makebox(0,0)[r]{ 0.02}}
\put(1419.0,270.0){\rule[-0.200pt]{4.818pt}{0.400pt}}
\put(201.0,418.0){\rule[-0.200pt]{4.818pt}{0.400pt}}
\put(181,418){\makebox(0,0)[r]{ 0.04}}
\put(1419.0,418.0){\rule[-0.200pt]{4.818pt}{0.400pt}}
\put(201.0,565.0){\rule[-0.200pt]{4.818pt}{0.400pt}}
\put(181,565){\makebox(0,0)[r]{ 0.06}}
\put(1419.0,565.0){\rule[-0.200pt]{4.818pt}{0.400pt}}
\put(201.0,713.0){\rule[-0.200pt]{4.818pt}{0.400pt}}
\put(181,713){\makebox(0,0)[r]{ 0.08}}
\put(1419.0,713.0){\rule[-0.200pt]{4.818pt}{0.400pt}}
\put(201.0,860.0){\rule[-0.200pt]{4.818pt}{0.400pt}}
\put(181,860){\makebox(0,0)[r]{ 0.1}}
\put(1419.0,860.0){\rule[-0.200pt]{4.818pt}{0.400pt}}
\put(201.0,123.0){\rule[-0.200pt]{0.400pt}{4.818pt}}
\put(201,82){\makebox(0,0){ 0.15}}
\put(201.0,840.0){\rule[-0.200pt]{0.400pt}{4.818pt}}
\put(449.0,123.0){\rule[-0.200pt]{0.400pt}{4.818pt}}
\put(449,82){\makebox(0,0){ 0.2}}
\put(449.0,840.0){\rule[-0.200pt]{0.400pt}{4.818pt}}
\put(696.0,123.0){\rule[-0.200pt]{0.400pt}{4.818pt}}
\put(696,82){\makebox(0,0){ 0.25}}
\put(696.0,840.0){\rule[-0.200pt]{0.400pt}{4.818pt}}
\put(944.0,123.0){\rule[-0.200pt]{0.400pt}{4.818pt}}
\put(944,82){\makebox(0,0){ 0.3}}
\put(944.0,840.0){\rule[-0.200pt]{0.400pt}{4.818pt}}
\put(1191.0,123.0){\rule[-0.200pt]{0.400pt}{4.818pt}}
\put(1191,82){\makebox(0,0){ 0.35}}
\put(1191.0,840.0){\rule[-0.200pt]{0.400pt}{4.818pt}}
\put(1439.0,123.0){\rule[-0.200pt]{0.400pt}{4.818pt}}
\put(1439,82){\makebox(0,0){ 0.4}}
\put(1439.0,840.0){\rule[-0.200pt]{0.400pt}{4.818pt}}
\put(201.0,123.0){\rule[-0.200pt]{298.234pt}{0.400pt}}
\put(1439.0,123.0){\rule[-0.200pt]{0.400pt}{177.543pt}}
\put(201.0,860.0){\rule[-0.200pt]{298.234pt}{0.400pt}}
\put(40,491){\makebox(0,0){$b$}}
\put(820,21){\makebox(0,0){$\Omega_m$}}

\put(690,527){\makebox(0,0)[l]{68\% CL}}
\put(690,387){\makebox(0,0)[l]{90\% CL }}
\put(944,786){\makebox(0,0)[l]{ SNe Ia + BAO + CMB}}
\put(201.0,123.0){\rule[-0.200pt]{0.400pt}{177.543pt}}
\put(627,576){\raisebox{-.8pt}{\makebox(0,0){$\cdot$}}}
\put(627,652){\raisebox{-.8pt}{\makebox(0,0){$\cdot$}}}
\put(637,555){\raisebox{-.8pt}{\makebox(0,0){$\cdot$}}}
\put(637,674){\raisebox{-.8pt}{\makebox(0,0){$\cdot$}}}
\put(647,541){\raisebox{-.8pt}{\makebox(0,0){$\cdot$}}}
\put(647,689){\raisebox{-.8pt}{\makebox(0,0){$\cdot$}}}
\put(657,530){\raisebox{-.8pt}{\makebox(0,0){$\cdot$}}}
\put(657,701){\raisebox{-.8pt}{\makebox(0,0){$\cdot$}}}
\put(666,521){\raisebox{-.8pt}{\makebox(0,0){$\cdot$}}}
\put(666,710){\raisebox{-.8pt}{\makebox(0,0){$\cdot$}}}
\put(676,514){\raisebox{-.8pt}{\makebox(0,0){$\cdot$}}}
\put(676,717){\raisebox{-.8pt}{\makebox(0,0){$\cdot$}}}
\put(686,508){\raisebox{-.8pt}{\makebox(0,0){$\cdot$}}}
\put(686,724){\raisebox{-.8pt}{\makebox(0,0){$\cdot$}}}
\put(696,503){\raisebox{-.8pt}{\makebox(0,0){$\cdot$}}}
\put(696,729){\raisebox{-.8pt}{\makebox(0,0){$\cdot$}}}
\put(706,499){\raisebox{-.8pt}{\makebox(0,0){$\cdot$}}}
\put(706,734){\raisebox{-.8pt}{\makebox(0,0){$\cdot$}}}
\put(716,495){\raisebox{-.8pt}{\makebox(0,0){$\cdot$}}}
\put(716,737){\raisebox{-.8pt}{\makebox(0,0){$\cdot$}}}
\put(726,492){\raisebox{-.8pt}{\makebox(0,0){$\cdot$}}}
\put(726,740){\raisebox{-.8pt}{\makebox(0,0){$\cdot$}}}
\put(736,490){\raisebox{-.8pt}{\makebox(0,0){$\cdot$}}}
\put(736,742){\raisebox{-.8pt}{\makebox(0,0){$\cdot$}}}
\put(746,488){\raisebox{-.8pt}{\makebox(0,0){$\cdot$}}}
\put(746,744){\raisebox{-.8pt}{\makebox(0,0){$\cdot$}}}
\put(756,487){\raisebox{-.8pt}{\makebox(0,0){$\cdot$}}}
\put(756,745){\raisebox{-.8pt}{\makebox(0,0){$\cdot$}}}
\put(766,486){\raisebox{-.8pt}{\makebox(0,0){$\cdot$}}}
\put(766,746){\raisebox{-.8pt}{\makebox(0,0){$\cdot$}}}
\put(775,486){\raisebox{-.8pt}{\makebox(0,0){$\cdot$}}}
\put(775,746){\raisebox{-.8pt}{\makebox(0,0){$\cdot$}}}
\put(785,486){\raisebox{-.8pt}{\makebox(0,0){$\cdot$}}}
\put(785,745){\raisebox{-.8pt}{\makebox(0,0){$\cdot$}}}
\put(795,487){\raisebox{-.8pt}{\makebox(0,0){$\cdot$}}}
\put(795,744){\raisebox{-.8pt}{\makebox(0,0){$\cdot$}}}
\put(805,488){\raisebox{-.8pt}{\makebox(0,0){$\cdot$}}}
\put(805,743){\raisebox{-.8pt}{\makebox(0,0){$\cdot$}}}
\put(815,490){\raisebox{-.8pt}{\makebox(0,0){$\cdot$}}}
\put(815,741){\raisebox{-.8pt}{\makebox(0,0){$\cdot$}}}
\put(825,492){\raisebox{-.8pt}{\makebox(0,0){$\cdot$}}}
\put(825,738){\raisebox{-.8pt}{\makebox(0,0){$\cdot$}}}
\put(835,495){\raisebox{-.8pt}{\makebox(0,0){$\cdot$}}}
\put(835,735){\raisebox{-.8pt}{\makebox(0,0){$\cdot$}}}
\put(845,498){\raisebox{-.8pt}{\makebox(0,0){$\cdot$}}}
\put(845,731){\raisebox{-.8pt}{\makebox(0,0){$\cdot$}}}
\put(855,501){\raisebox{-.8pt}{\makebox(0,0){$\cdot$}}}
\put(855,727){\raisebox{-.8pt}{\makebox(0,0){$\cdot$}}}
\put(865,506){\raisebox{-.8pt}{\makebox(0,0){$\cdot$}}}
\put(865,722){\raisebox{-.8pt}{\makebox(0,0){$\cdot$}}}
\put(874,511){\raisebox{-.8pt}{\makebox(0,0){$\cdot$}}}
\put(874,716){\raisebox{-.8pt}{\makebox(0,0){$\cdot$}}}
\put(884,517){\raisebox{-.8pt}{\makebox(0,0){$\cdot$}}}
\put(884,709){\raisebox{-.8pt}{\makebox(0,0){$\cdot$}}}
\put(894,524){\raisebox{-.8pt}{\makebox(0,0){$\cdot$}}}
\put(894,702){\raisebox{-.8pt}{\makebox(0,0){$\cdot$}}}
\put(904,532){\raisebox{-.8pt}{\makebox(0,0){$\cdot$}}}
\put(904,693){\raisebox{-.8pt}{\makebox(0,0){$\cdot$}}}
\put(914,542){\raisebox{-.8pt}{\makebox(0,0){$\cdot$}}}
\put(914,682){\raisebox{-.8pt}{\makebox(0,0){$\cdot$}}}
\put(924,554){\raisebox{-.8pt}{\makebox(0,0){$\cdot$}}}
\put(924,668){\raisebox{-.8pt}{\makebox(0,0){$\cdot$}}}
\put(934,572){\raisebox{-.8pt}{\makebox(0,0){$\cdot$}}}
\put(934,650){\raisebox{-.8pt}{\makebox(0,0){$\cdot$}}}
\put(567,570){\makebox(0,0){$\cdot$}}
\put(567,660){\makebox(0,0){$\cdot$}}
\put(577,544){\makebox(0,0){$\cdot$}}
\put(577,688){\makebox(0,0){$\cdot$}}
\put(587,526){\makebox(0,0){$\cdot$}}
\put(587,707){\makebox(0,0){$\cdot$}}
\put(597,512){\makebox(0,0){$\cdot$}}
\put(597,722){\makebox(0,0){$\cdot$}}
\put(607,501){\makebox(0,0){$\cdot$}}
\put(607,734){\makebox(0,0){$\cdot$}}
\put(617,491){\makebox(0,0){$\cdot$}}
\put(617,744){\makebox(0,0){$\cdot$}}
\put(627,483){\makebox(0,0){$\cdot$}}
\put(627,753){\makebox(0,0){$\cdot$}}
\put(637,475){\makebox(0,0){$\cdot$}}
\put(637,761){\makebox(0,0){$\cdot$}}
\put(647,469){\makebox(0,0){$\cdot$}}
\put(647,768){\makebox(0,0){$\cdot$}}
\put(657,464){\makebox(0,0){$\cdot$}}
\put(657,774){\makebox(0,0){$\cdot$}}
\put(666,459){\makebox(0,0){$\cdot$}}
\put(666,780){\makebox(0,0){$\cdot$}}
\put(676,454){\makebox(0,0){$\cdot$}}
\put(676,784){\makebox(0,0){$\cdot$}}
\put(686,451){\makebox(0,0){$\cdot$}}
\put(686,788){\makebox(0,0){$\cdot$}}
\put(696,447){\makebox(0,0){$\cdot$}}
\put(696,792){\makebox(0,0){$\cdot$}}
\put(706,445){\makebox(0,0){$\cdot$}}
\put(706,795){\makebox(0,0){$\cdot$}}
\put(716,442){\makebox(0,0){$\cdot$}}
\put(716,797){\makebox(0,0){$\cdot$}}
\put(726,440){\makebox(0,0){$\cdot$}}
\put(726,799){\makebox(0,0){$\cdot$}}
\put(736,439){\makebox(0,0){$\cdot$}}
\put(736,801){\makebox(0,0){$\cdot$}}
\put(746,438){\makebox(0,0){$\cdot$}}
\put(746,802){\makebox(0,0){$\cdot$}}
\put(756,437){\makebox(0,0){$\cdot$}}
\put(756,803){\makebox(0,0){$\cdot$}}
\put(766,436){\makebox(0,0){$\cdot$}}
\put(766,803){\makebox(0,0){$\cdot$}}
\put(775,436){\makebox(0,0){$\cdot$}}
\put(775,803){\makebox(0,0){$\cdot$}}
\put(785,436){\makebox(0,0){$\cdot$}}
\put(785,803){\makebox(0,0){$\cdot$}}
\put(795,436){\makebox(0,0){$\cdot$}}
\put(795,802){\makebox(0,0){$\cdot$}}
\put(805,437){\makebox(0,0){$\cdot$}}
\put(805,801){\makebox(0,0){$\cdot$}}
\put(815,438){\makebox(0,0){$\cdot$}}
\put(815,800){\makebox(0,0){$\cdot$}}
\put(825,440){\makebox(0,0){$\cdot$}}
\put(825,798){\makebox(0,0){$\cdot$}}
\put(835,441){\makebox(0,0){$\cdot$}}
\put(835,795){\makebox(0,0){$\cdot$}}
\put(845,443){\makebox(0,0){$\cdot$}}
\put(845,793){\makebox(0,0){$\cdot$}}
\put(855,446){\makebox(0,0){$\cdot$}}
\put(855,790){\makebox(0,0){$\cdot$}}
\put(865,448){\makebox(0,0){$\cdot$}}
\put(865,787){\makebox(0,0){$\cdot$}}
\put(874,451){\makebox(0,0){$\cdot$}}
\put(874,783){\makebox(0,0){$\cdot$}}
\put(884,455){\makebox(0,0){$\cdot$}}
\put(884,779){\makebox(0,0){$\cdot$}}
\put(894,459){\makebox(0,0){$\cdot$}}
\put(894,774){\makebox(0,0){$\cdot$}}
\put(904,463){\makebox(0,0){$\cdot$}}
\put(904,769){\makebox(0,0){$\cdot$}}
\put(914,467){\makebox(0,0){$\cdot$}}
\put(914,763){\makebox(0,0){$\cdot$}}
\put(924,473){\makebox(0,0){$\cdot$}}
\put(924,757){\makebox(0,0){$\cdot$}}
\put(934,479){\makebox(0,0){$\cdot$}}
\put(934,750){\makebox(0,0){$\cdot$}}
\put(944,485){\makebox(0,0){$\cdot$}}
\put(944,743){\makebox(0,0){$\cdot$}}
\put(954,492){\makebox(0,0){$\cdot$}}
\put(954,734){\makebox(0,0){$\cdot$}}
\put(964,500){\makebox(0,0){$\cdot$}}
\put(964,725){\makebox(0,0){$\cdot$}}
\put(974,510){\makebox(0,0){$\cdot$}}
\put(974,714){\makebox(0,0){$\cdot$}}
\put(983,521){\makebox(0,0){$\cdot$}}
\put(983,702){\makebox(0,0){$\cdot$}}
\put(993,534){\makebox(0,0){$\cdot$}}
\put(993,688){\makebox(0,0){$\cdot$}}
\put(1003,550){\makebox(0,0){$\cdot$}}
\put(1003,670){\makebox(0,0){$\cdot$}}
\put(1013,576){\makebox(0,0){$\cdot$}}
\put(1013,643){\makebox(0,0){$\cdot$}}
\end{picture}
\input{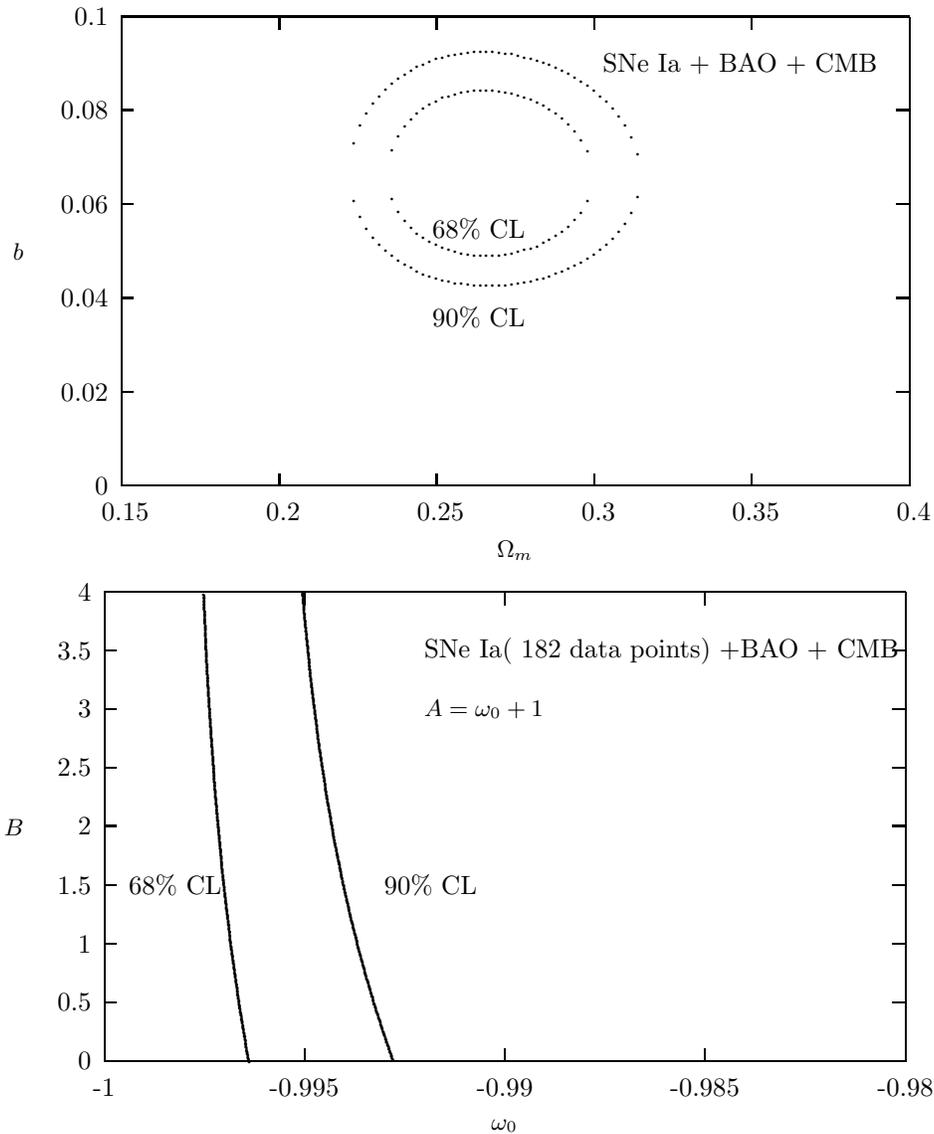}}
\caption{{\bf{a)}} Confidence regions in the $b - \Omega_m$ plane (Parameterization P1) arising from SNe Ia (\emph{gold} sample with 182 points) $+$ CMB (WMAP 3 yr) $+$ SDSS BAO measurement. {\bf{b)}} Confidence contours in the plane $B- \omega_0$ for P2.}.
\end{figure*}

\section{Results}

For the parameterization P1, the cosmological parameters are $\Omega_m$ and $b$. Figure 1 shows the contours corresponding to $ 68 \% $  and $90 \%$ CL in the parametric space for the undulant universe. From the combined analysis discussed in the section III, we calculate  the best fit value for the parameters, i.e., $ b = 0.06 \pm 0.01$ and   $\Omega_m = 0.26\pm 0.03 $ at $ 1\sigma$ level. These limits are in agreement with earlier results showing that the undulant parameterizations can reproduce the CMB temperature asymmetries and the correct power spectrum if the frequency parameter $ b\le 2$ \cite{un}, as well as with current X-ray  gas mass fraction measurements of dynamically relaxed galaxy clusters that imply $ b\ge 2.2$. The linear treatment of the evolution of the mass fluctuations also restricts the frequency parameter to $b \le 0.4$ \cite{un}. Note that,  although in full agreement with the previously estimates of the parameter $b$ \cite{un},  the constraints found here are much tighter.

From  Eq. (1), one can easily see that in the limit $b \rightarrow 0$, the model P1 behave as cosmological constant. So with this small value of  $b$,  one can conclude the undulant model  is behaving like a cosmological constant which is in agreement with the present cosmological observations.

For parameterization P2, we work with the  condition that $\omega (z) \ge -1$ (Quintessence field) which imply $ \omega_0 - A = -1$. We marginalize over $\Omega_m$ assuming the gaussian prior  $\Omega_m = 0.234 \pm 0.035$ provided by WMAP 3 year data \cite{cmb2}. Thus the remaining parameters  are $B$ and $\omega_0$.  Figure 2 shows the  contours corresponding to $ 68 \% $  and $90 \%$ CL in the parametric space. In this model the entire range of $B$ is allowed. However, the constraints on $\omega_0$ are very tight: $\omega_0 <-0.996$ at $1\sigma$ and $\omega_0 < -0.992$ at $ 90 \%$  confidence level. The best fit values for this model is $\omega_0 = -1.0 $ which gives $A = 0$. Even at $1 \sigma$ and at $ 90 \%$  confidence level the value of  $\omega_0$ is close to $- 1$, hence  the amplitude of oscillations, $A$, approaches to  zero. Therefore P2 also behaves like  the $\Lambda$CDM model. Note also that, since the entire range of $B$ is  allowed at $2 \sigma$ level, this is in complete agreement with the results obtained earlier in Ref. \cite{lin} in which it is shown that for small amplitudes any
value of frequency B is allowed by distance data.

\section{Final remarks and discussion}

A possible oscilatory evolution of the Universe seems to have some support from current observational data as, for instance, (i) the north-south pencil beam survey that pointed out the periodicity observed in the galaxy redshift distribution at the regular intervals of 128 Mpc \cite{p}. This periodic distribution could be in principle explained by the oscillating expansion \cite{o}; the comparative study of various dark energy parameterizations with \emph{gold} SNe Ia data that have shown that the best fit model is achieved by oscillating parameterizations \cite{peri}; (iii) the multipole moments $C_l$ in the CMB spectra that can be improved if  we superimposed the oscillations in the first year WMAP data \cite{cm}.  

In this work we have discussed observational constraints on the two purely phenomenological models for $w(z)$ which describes oscillating scenarios. To this end we have used the CMB shift parameter derived from the recent WMAP 3 year data together with  SNe Ia data (\emph{gold} sample) and the LSS data (BAO measurement from the  SDSS). We find that both models behave like the standard $\Lambda$CDM scenario, which confirms the conclusion that we will have to wait for the next generation observations to extract information about the evolution of $\omega$ with $z$.  Only a complete and coherent theory of dark energy selected by the observations can predict the fate of the universe correctly. This certainly will require high level of precision and  control of systematics in observations. 



\begin{section}*{Acknowledgments}
We are thankful to E. V. Linder and  G. Barenboim for useful discussions. The authors (D. Jain $\&$ A. Dev) also thank Amitabha Mukherjee and Shobhit Mahajan for providing  the facilities to carry out the research work.  J.S. Alcaniz is supported by CNPq (No. 307860/2004-3 and  No. 475835/2004-2) and by FAPERJ (No. E-26/171.251/2004).
\end{section}

\end{document}